# Online Deliberation Design: Choices, Criteria, and Evidence

Todd Davies and Reid Chandler
Symbolic Systems Program, Stanford University

*Abstract:* This chapter reviews empirical evidence bearing on the design of online forums for deliberative civic engagement. Dimensions of design are defined for different aspects of the deliberation: its purpose, the target population, the spatiotemporal distance separating participants, the communication medium, and the deliberative process to be followed. After a brief overview of criteria for evaluating different design options, empirical findings are organized around design choices. Research has evolved away from treating technology for online deliberation dichotomously (either present or not) toward nuanced findings that differentiate between technological features, ways of using them, and cultural settings. The effectiveness of online deliberation depends on how well the communicative environment is matched to the deliberative task. Tradeoffs, e.g. between rich and lean media and between anonymous and identifiable participation, suggest different designs depending on the purpose and participants. Findings are limited by existing technologies, and may change as technologies and users co-evolve.

"Creating democracy is ... harder than rocket science." - Ira Harkavy

## DESIGNING FOR DELIBERATION

As they evolve, information and communication technologies are providing numerous and growing alternative ways for people to interact with each other and with information. In turn, this evolution is providing more choices in the creation of online deliberative civic engagement forums, sometimes called deliberative e-democracy. This chapter focuses on the design of deliberative e-democracy forums in the light of both available and prospective information/communication technologies. The space of possibilities is vast, so we will touch on just some of the more significant choices that must be made, as well as the current state of empirical research bearing on these choices.

We will take "deliberation" to denote "thoughtful, careful, or lengthy consideration" by individuals, and "formal discussion and debate" in groups.[1] We are therefore primarily interested in communication that is reasoned, purposeful, and interactive (for a discussion about communication in deliberative civic engagement, see chapter 4). However, the power and predominance of other influences on political decisions and communication (e.g., mass media, appeals to emotion and authority, and snap judgments) are obviously also relevant to deliberative e-democracy, both as contrasts to deliberative discourse and as potential influences on deliberation.

The term "online" as a modifier to "deliberation" could be read to indicate the mediation of deliberation among participants through one or more electronic communication technologies that augment our usual abilities to see or hear information separated from us in time or space. In addition to the Internet, this would include telephone (including "smart phone") and teleconferencing systems, broadcasting (if used to facilitate communication between participants, e.g., "talking heads" debating over different satellite feeds, rather than just presenting information relevant to deliberators), and electronic tools through which participants in face-to-face meetings interact.

**Design Choices**

Once a decision is made to have online deliberation, the creators and conveners need to make several design choices. Below we discuss a set of broad design issues, grouped into categories. These issues have been selected for inclusion mainly because they have been the subjects of empirical research. As should be apparent, the issues discussed below represent a relatively small subset of all the potential choices a deliberation designer faces. Nevertheless, these issues do provide a way to organize the empirical literature relevant to deliberation design.[2] The design choices are grouped under five categories representing the highest level questions faced by the deliberation designer: 1) Purpose (*Why* is the deliberation being designed – in other words, what objectives should the design reflect?); 2) Population (*Who* will be involved?); 3) Spatiotemporal Distance (*Where* and *when* will participants be interacting with each other?); 4) Communication Medium (*How* will communication occur?); and 5) Deliberative Process (*What* will occur among the participants?). These categories, and all of the choice dimensions described within them (see Table 6.1 for a summary), could be applied both to offline (i.e., face-to-face) and online deliberation. Our focus in what follows, however, will be on how the availability of communication technologies affects these choices. We describe each dimension more fully below. In the second section of the chapter, we discuss empirical findings related to each dimension. We then conclude with a discussion of the main lessons that emerge from this empirical literature.[3]

---

[1] "From Latin deliberare to consider well" (Collins English Dictionary 1979). "Deliberation" can also refer to thinking processes within an individual mind. For this article, however, we focus on deliberation as a type of communication between people. Extensive definitions of deliberation have been proposed in the literature, e.g., Burkhalter, Gastil, and Kelshaw (2002) who focus on criteria for democratic deliberation in small face-to-face groups, as well as the literature on coding for deliberativeness or deliberative quality, in which measures could be said to constitute a definition of deliberation (see Graham and Witschge 2003; Stromer-Galley 2007; Trénel 2004).

[2] For other dimensional analyses of communication forms in general, see Clark (1996), Eckles, Ballagas, and Takayama (2009).

[3] For some other typologies of issues in the design of online deliberation systems, see Rose and Sæbø (2010), and De Cindio and Peraboni (2010).

**Table 6.1: Summary of Five Design Categories**

| Category | Question | Design Dimensions |
|---|---|---|
| *1. Purpose* | *Why* is the deliberation being designed? | (a) Outcome (decisions-beliefs-ideas) <br> (b) Collectivity (group-individual) |
| *2. Population* | *Who* will be involved? | (a) Recruitment (random-selected) <br> (b) Audience (public-private) |
| *3. Spatiotemporal Distance* | *Where* and *when* will participants be interacting with each other? | (a) Colocation (face to face-telecommunication) <br> (b) Cotemporality (synchronous-asynchronous) |
| *4. Communication Medium* | *How* will communication occur? | (a) Modality (speech-text-image-multimodal) <br> (b) Emotivity (impeded-enabled) <br> (c) Fidelity (transformed-unaltered) |
| *5. Deliberative Process* | *What* will occur between participants | (a) Facilitation (moderated-unmoderated) <br> (b) Structure (rules-free form) <br> (c) Identifiability (identifiable-anonymous) <br> (d) Incentivization (reward-no reward) |

*Purpose*

The purposive nature of deliberation is represented in our design space with two dimensions: the outcome of the deliberation and its collectivity.

*Outcome*. Should the goal of the deliberation be to produce a decision, to affect opinions or knowledge, or to generate ideas? While both decision-making and brainstorming are legitimate goals/purposes of deliberation, it is generally easier to determine when deliberation has reached its end state when one or more decisions are to be made. Deliberations undertaken in order to make decisions also tend to be different in character from those whose aim is idea generation, with the former being more oriented toward debating options that compete with each other and the latter fostering sustained openness to different ideas and even an appreciation for their variety. Decision-making and belief formation are generally more demanding than brainstorming in an online environment, because they are somewhat more likely to require access to tools (such as calculation aids), as well as information that participants may expect to be able to consult before making a decision.

*Collectivity*. Should the product/goal of deliberation be an outcome for a whole group, or just for individuals within a deliberating group? Whether the outcome is a set of decisions, beliefs, or ideas is independent of whether those decisions/beliefs/ideas are associated with the group collectively or only with individual participants (or subgroups). The collectivity dimension, like the outcome dimension, has deep implications for a deliberation's design. If the deliberating group must agree collectively on a set of decisions or ideas, more pressure to achieve mutual understanding is put on the process, and it will be harder to end the deliberation at an arbitrary point, such as when a time limit has been reached. Collective/agreed outcomes pose special challenges for an online environment, because the tools required to reach consensus or to take a vote are more elaborate than a simple threaded conversation, which might be sufficient for brainstorming or for idea generation.

*Population*

A consideration of the first question above (the purpose of deliberation) obviously entails some assumptions about whom the deliberation is for and who will participate (for a more detailed discussion, see chapter 3). But the details are probably best considered only after the purpose of a deliberation has been determined. In our framework, population choices are represented by two dimensions: the recruitment of participants and the audience for the deliberative exercise.

*Recruitment.* Will participants represent a random sample, a self-selected group, or some other sampling criterion? The question of recruitment looms large for both offline and online deliberations, especially for public deliberations that are meant to inform public policy. When information/communication technologies are involved, however, there are special considerations. Since not everyone has ready access to the Internet, for example, achieving a representative sample poses extra challenges for online deliberation. Still, participants in online public deliberations can be and have been given computers and Internet access in exchange for their participation, and other techniques such as adjusted sampling can be used to achieve representativeness when participant demographics are known.[4]

*Audience.* Will the deliberation process be visible to the public (either during or at some point after deliberation), or will it remain private to the deliberating group?[5] Even when the issue to be deliberated upon is a matter of public concern, the deliberation may be set up in a way that maintains confidentiality for the participants with respect to the larger public. A televised broadcast of a deliberation would not qualify under the definition of "online" given above unless the broadcast itself is integral to the deliberation (e.g., if viewers can be actively involved). But a broadcast deliberation can, in any case, greatly expand the audience for a deliberation, and communication technologies generally have the potential to make either passive or active participation much more widespread. The use of online tools has implications for the eventual audience, since online deliberation can generally be recorded, and it is difficult or impossible for participants to rely on assurances that their participation will not be shared with others when deliberation occurs online.

---

[4] See for example, Fishkin (2009c), Price (2009), and Cavalier (2009).

[5] Eckles, Ballagas, and Takayama (2009: 14-21) distinguish the audience of an act of communication from the "addressee", who may be an individual, a group, or a general audience. This is an example of further refinement that is possible in defining the design space for deliberation, but since the present volume is concerned with public deliberation, we felt it was outside the scope of this chapter to include cases where communication occurs between individuals privately. It should be noted, however, that a designer of a public deliberation system might want to allow for "backchannel" communication between individual participants. Thus, analyses like those of Eckles, Ballagas, and Takayama (2009) are relevant to the design of online deliberation systems at a more detailed level than we discuss here.

*Spatiotemporal Distance*

Information/communication technologies greatly expand the options for participants in a deliberation to be separated from each other in space and/or time. Our framework analyzes these possibilities into two choice dimensions: the colocation and the cotemporality of participants.

*Colocation*. Will the deliberation happen face-to-face (at the same time and place) or via some form of space- or time-shifted communication? "Location" here refers to a point or region in both space and time. If all participants are present in the same place at the same time, we will call this a face-to-face deliberation, even though the group might be sufficiently large so that not all participants can see each other. When some or all participants are spatially separated from each other, some form of telecommunication[6] technology is necessary. In either case, deliberation can happen "online" in the sense described above. For example, in a face-to-face deliberation, a shared display can be used to record the proceedings for all to see, as it is happening. Participants can also interact with each other via devices that record survey responses displayed in real time, or via a chatroom that is accessed on individuals' laptops during a meeting. The use of these technologies in a face-to-face event might be somewhat distracting, but generally they do not remove the affordances of sharing space with other participants and can usefully augment face-to-face dialogue. In telecommunication deliberations, on the other hand, such as a teleconference, technology can be used that simulates face-to-face interaction to a greater or lesser extent. Colocation should probably therefore be viewed as a set of continua, depending on whether some, most, or all participants are colocated, and to what extent participants' behavior and subjective feeling during deliberation reflects colocation, regardless of whether they are physically co-present.[7]

*Cotemporality*. Will participation be synchronous or asynchronous? Once a decision has been made to use telecommunication, a forum designer must decide how much if any communication will be allowed to happen at some temporal distance, viz asynchronously. Asynchrony allows participation to occur at different times for different people, as happens for example in email or a message board. Synchronous or cotemporal participation implies that communication is received as soon as it is emitted. A face-to-face meeting is generally synchronous, but so is real-time text chat. Thus, cotemporality does not imply much about the medium or modality of communication. Asynchronous communication has traditionally been associated more with text than with speech, but technology has closed this gap. The advent of voicemail, for example, turned the telephone from a strictly synchronous medium into an asynchronous one as well. Some media afford both synchronous and asynchronous communication, depending on the participants. Text messaging, for example, can be considered a hybrid of the two, because a recipient of a text message can read and respond to it right away, or later. It might be termed "on demand" cotemporality. Thus, cotemporality, like colocation, should be seen as a continuum of possibilities.[8] Actual deliberation events may also combine synchronous and asynchronous communication, for example as when ideas from each table at an America*Speaks*' deliberation are submitted via laptops to "synthesizers" who combine and post ideas on a central display.[9]

---

[6] "Telecommunication" means communication "over a distance..." (Collins English Dictionary 1979). We include "temporal distance" in our use of this concept, because the technologies that allow communication across time at the same location appear to be the same as those that allow it across space and time combined.

[7] Clark and Brennan (1991) and Eckles, Ballagas, and Takayama (2009: 10) define a purely spatial dimension, which allows for the case of participants being in the same spatial location but separated in time. A stationary bulletin board or kiosk could facilitate deliberation in such a setting. We choose to treat this case together with space-shifted communication within our framework, but note that there is this further refinement that could be introduced if necessary.

[8] Eckles, Ballagas, and Takayama (2009: 11) make a finer distinction here, between a "record" dimension (whether communication is ephemeral or available at an arbitrary time later) and a "synchronicity" dimension (whether a reply can be delayed or not).

### *Communication Medium*

A deliberation designer must make detailed choices regarding the media through which deliberation will occur. Consistent with the rest of our framework, we consider just a few variables (dimensions) characterizing different media. This way of presenting the design space for communication media has been called "morphological analysis."[10] It is also in the tradition of "variable-based approaches" to the empirical study of communication technology.[11] We characterize communication media by their modalities, emotivity, and fidelity.

*Modalities*. Should communication occur through voice, text, images, or in a multimodal way? The modality of a communication is defined by the perceptual and cognitive faculties it engages, in contrast to the communication medium, which is the technology that delivers it. Here the range of design choices is truly vast. Participants can receive information through text (reading) and produce through voice (speaking), or they can listen and write. They can be given a choice of several modalities, or be restricted to one. Prepared background information (e.g., materials provided prior to a deliberation) can be transmitted one way, while communication between participants can happen via a different modality. Information can be presented in multiple modalities (e.g., as audio and video) or as text with pictures. Again, technology keeps enhancing the set of possibilities. Speech recognition systems, for example, are beginning to offer the possibility of a textual transcript in real time as people speak.[12] Many technologies now make it increasingly possible to illustrate what one is saying or writing pictorially or through video. Thus, the choice of modality is becoming truly independent of other choices, such as whether to colocate or synchronize a deliberation.

*Emotivity*. Will the medium encourage emotional involvement and/or expression, and if so, what kinds? In theory, the term "deliberation" connotes calm, relatively unemotional reasoning; but in practice, deliberation often concerns deeply emotional issues (see chapter 4 for a discussion). People's contributions or the results of deliberation may be enhanced if they find the experience fun, or at least pleasant. Or, they may learn more from encounters that are unsettling or even sad or angering. Some researchers have argued that better decisions correspond to less emotional ones.[13] The place of emotion is a question shared between online and offline deliberation design, but online environments introduce their own issues related to emotion. Telecommunication can strip out emotionality, particularly if emotional cues (such as facial expressions and body language) are absent. On the other hand, the isolation of a participant on the Internet, for example, might lead them to feel and/or express more emotion than they would face-to-face. Online forums sometimes provide features such as emoticons to enable emotional expression. Both offline and online, then, whether and how much to encourage/enable emotional expression is an important design decision.

*Fidelity*. Will participants' communicative acts be transformed or left unaltered by the medium? This is a somewhat exotic dimension that seems likely to grow in importance in the future. "Transformed social interaction" (TSI) is communication in which some aspect of a communicator's message is systematically altered in order to change the experience of the recipient.[14] Virtual reality systems (for example, Second Life) in which participants are represented by avatars are an example of transformed communication. Moreover, in such virtual reality systems, it is possible for all participants to make (virtual) eye contact with each other at the same time.[15] But transformation can and does occur without electronic technology as well. For example, when a newspaper editor alters the text of a submitted article or letter prior to publishing it, the result is transformed communication. Newspapers, then, can be a transformational medium. Technology has the potential to dramatically increase the scope and sophistication of transformed communication. For example, experiments are being conducted to simulate future artificial intelligence devices that correct one's grammar or word choices in mid-communication.[16]

---

[9] See Lukensmeyer, Goldman, and Brigham (2005). Thanks to Peter Muhlberger for pointing out this possibility.

[10] Zwicky (1967) and Card, Mackinlay, and Robertson (1991), both cited in Eckles, Barragas, and Takayama (2009: 4).

[11] Nass and Mason (1990), cited in Eckles, Barragas, and Takayama (2009: 4).

[12] See Anusuya and Katti (2009).

[13] See Greene (2003).

### *Deliberative Process*

Finally, and perhaps most importantly, the designer of a deliberation must decide what process will be followed during the deliberation. Of course, the participants themselves could decide this, and they sometimes insist on doing so (as they may for the other design categories noted above) even if the designer has other plans. However, planning for participants to design all or part of the process is still a choice for the designer. Our framework includes four choice dimensions concerning the deliberative process: facilitation, structure, identifiability, and incentivization.

*Facilitation*. Will the deliberation be facilitated by a moderator, and if so how? Facilitation can occur in many different forms, including no facilitation at all. If one or more facilitators or moderators are present, they can take on many different roles. One researcher offers the following list of possibilities:

1. Greeter: making people feel welcome;
2. Conversation Stimulator: posing new questions and topics, playing devil's advocate in existing conversation;
3. Conflict Resolver: mediating conflicts towards collective agreements (or agreeing to disagree);
4. Summarizer of Debates[17];
5. Problem Solver: directing questions to relevant people for response;
6. Supporter: bringing in external information to enrich debates and support arguments;
7. Welcomer: bringing in new participants, either citizens or politicians/civil servants;
8. 'Cybrarian': providing expert knowledge on particular topics;

---

14 Bailenson, Beall, Loomis, Blascovich, and Turk (2004).

15 Bailenson, Beall, Loomis, Blascovich, and Turk (2005: 511).

16 Yamada, Nakajima, Lee, Brave, Maldonado, Nas, and Morishima (2008).

17 For example, the "synthesizers" of discussions at town hall meetings. [This note is not in the original quotation.]

9. Open Censor: deleting messages deemed inappropriate, normally against predefined rules and criteria. Feedback is given to explain why, and an opportunity to rewrite is provided;
10. Covert Censor: deleting messages deemed inappropriate, but without explaining why;
11. Cleaner: removing or closing dead threads, hiving off sub-discussions into separate threads. [18]

A facilitator can take on any subset of these roles, making the space of possibilities quite large. In this chapter, we emphasize the online version of facilitation, but obviously many issues are shared with offline deliberation. Some online environments afford the moderator a great deal of control that is not usually present in face-to-face deliberations, such as the power to censor comments before other participants are exposed to them. Additionally, advances in software make it possible to automate facilitation to a greater or lesser extent. Environments such as Slashdot.org contain built-in tools that aid people in classifying and giving provisional ratings to users' postings, which can both reflect and influence a user's reputation. Slashdot also recruits site members to assist in moderation, through simple feedback that is aggregated across users.

*Structure*. Will the deliberation be structured according to certain rules, or an organized progression, or will it be left unstructured? Online environments, in some ways, make it easier to structure deliberation, as technology can be a strong enforcer. In practice, however, imposing structure online can be more difficult than in a face-to-face meeting, especially if there is no moderator. We separate the issue of structure from facilitation, however, because structure can also exist in software or in the minds of participants (e.g., as a social norm) without a moderator being present.

*Identifiability*. Will participants be identifiable or anonymous? When each act of communication can be tied to an individual, participation is identifiable. Many forums, however, offer at least the option of contributing anonymously. Anonymity can exist at different levels. For example, one's identity may be invisible to most or all participants, but visible to others, at least to those administering the forum. True anonymity is rare in an online space, due to the traceability of IP addresses and other mechanisms (login requirements, HTTP cookies, browser fingerprinting, etc.) for identifying users online. This may be one respect in which offline deliberation, especially when not done face-to-face, offers possibilities that are inherently difficult to achieve if participants are online, since the world of bricks and mortar at least sometimes affords anonymous communication (e.g. anonymous postings on a wall or graffiti).

*Incentivizing*. Will participants be rewarded/reinforced for contributing, or not? While it has not often been an issue in face-to-face deliberations, the creation of an explicit reward or points system is a much more common feature of online than of offline forums. Some systems award points toward one's reputation and/or ability to contribute and to have one's messages flagged -- positively or negatively.[19] Back-end computation makes the necessary bookkeeping much easier, and in principle, such techniques could be introduced into face-to-face deliberations as well.

---

18 Wright (2009: 236).

19 See for example, Slashdot.org.

**Design Criteria**

Of course, it is difficult to decide how to design a deliberation forum unless we know how to assess its success. Here, we provide a brief discussion of some of the criteria that have been applied in the past, especially by researchers evaluating different design choices (for a more detailed discussion, see chapter 10). Which criteria are used is, of course, another design choice to be considered.

*Quantity*

We can measure how much deliberation or participation is taking place in different ways, for example by counting the total number of contributions, the average number of comments per participant, the total and average lengths of comments, and the number and proportion of participants who contribute.[20]

*Quality*

The quality of deliberative activity is more subjective than quantity, but it too can be measured in various ways. Metrics of quality include those that can be determined directly from the transcript(s) of a deliberation, e.g., the average number of replies to each message, the distribution of contributions to a thread from participants with different viewpoints, and the length of replies by others to a message. Quality can also be assessed outside of the deliberation itself, for example by asking participants to say or demonstrate how well they understood others, or to evaluate the quality of the deliberation.[21] A useful concept in assessing quality is *grounding* – the extent to which someone who says something knows that they have been understood correctly by those on the receiving end, and vice versa.[22] This differs from comprehension because, for example, a participant could be understood without knowing that they have been understood. Grounding is often assessed by inspecting a record of a dialogue (e.g., a transcript or video recording) and looking for markers of common understanding, such as mutual head nods, "yups" or "uh-huhs", and other linguistic acknowledgements. A few researchers have attempted to develop metrics for deliberativeness,[23] but as yet, no consensus measure or set of measures has emerged.

*Inclusiveness*

Measures of inclusiveness are, for example, how closely the set of participants matches the demographics of a target population. This is referred to as "external inclusiveness."[24] Other measures are related to "internal inclusiveness," for example, how often those who are in the minority, demographically or ideologically, contribute relative to other participants once they are part of a deliberation (for a discussion of diversity and inclusion in deliberative civic engagement, see chapter 5).

---

20 See for example, Kelly, Fisher, and Smith (2009).

21 See for example, Stromer-Galley and Muhlberger (2009).

22 Clark and Brennan (1991).

23 See for example, Jankowski and van Os (2004), Trénel (2004), Sack, Kelly, and Dale (2009), and Stromer-Galley (2007). See Muhlberger and Stromer-Galley (2009) for an automated approach based on statistically examining concept-word linkages in text.

24 Trénel (2009).

*Preference*

We can also ask whether participants or other stakeholders prefer a given type of deliberation over another, or over not deliberating at all. This can be measured prior to the deliberation itself, as a prospective choice, or during or after the experience, for example on a rating scale.

*Efficiency*

This criterion concerns the extent to which resources are optimally used. One measure is the speed with which deliberators can accomplish a task when time is allowed to vary. Other measures include cost and the number of people involved. The concept of "optimal" implies that this should be measured relative to the accomplishment of a goal, which may be harder to do.

*Efficacy*

"Deliberative efficacy" refers to the consequences of the deliberation – whether and to what extent the deliberation accomplishes its purposes. If the purpose is to foster mutual understanding, then measures such as grounding and retention are good metrics for efficacy. Participants may understand each other but nevertheless fail to agree or come to a resolution, for example on a decision or action. If this is the goal, efficacy must take into account whether the end result of the deliberation was consistent with that goal.

**SOME EMPIRICAL FINDINGS FROM THE LITERATURE**

Online deliberation is a relatively new field. Although the concept of public deliberation via electronic means was discussed as early as the 1970s,[25] and there was some early empirical work on deliberation online in the 1980s and 1990s,[26] studies of structured or public online deliberation appear to have begun with work by Stephen Coleman and colleagues,[27] Lincoln Dahlberg,[28] and Vincent Price[29] around a decade ago. A review of empirical literature relevant to online deliberation therefore benefits from forays into adjacent fields, such as the study of online communities, social software, computer supported cooperative work, human-computer interaction, psychology, education, information science, management science, political science, and media studies. What follows is a summary of some of the empirical work relevant to online deliberation, organized by the design choice dimensions defined in the previous section.

The focus in this chapter is on results that could not necessarily be known prior to empirical study. Much can be usefully said about the consequences of design choices that does not require such study. For example, online forums are easier and cheaper to set up for geographically dispersed participants, anonymity makes people less accountable for what they say, and Internet deliberation involves issues of accessibility and other digital divides. We have tried to limit ourselves to questions that require data analysis before we can answer them with confidence.

---

[25] See for example, Henderson (1970), Ohlin (1971), Etzioni (1972, 1975).
[26] See for example, Asteroff (1982), Schneider (1997), Wilhelm (1999).
[27] See for example, Blumler and Coleman (2001), Coleman, Hall, and Howell (2002).
[28] Dahlberg (2001).
[29] Price and Capella (2002); Price, Goldthwaite, and Capella (2002).

**Purpose**

*Outcome*

Whether a deliberation should be decision-, belief-, or idea-producing is partly determined by the context in which the need or desire for deliberation arises. However, there are some empirical results that bear on this question for deliberating groups. In comparisons of group versus individual brainstorming, the ideas generated have been found to be considerably less novel in group processes than when individuals are isolated,[30] because "deliberating groups discourage novelty."[31] Groups making decisions can similarly fail to reach good decisions when there is an initial bias among group members generally, or among its most influential members, away from the best outcome. In a typical study, business students were significantly more likely in groups than as individuals to invest more money in a failing project based on the "sunk cost fallacy."[32] The cognitive and motivational biases leading to "group polarization" – the tendency of groups to strengthen the average inclination of group members – can lead to *either* better or worse decisions after group discussion than in its absence.[33] When the right decision can be recognized easily by group members once it is identified (so-called "eureka problems"), group deliberation can outperform most if not all individuals within the group.[34] So these results, while they are not specific to the online case, suggest that deliberation has a better chance to improve decisions than to improve idea generation, but only if the decision is of the right type.

To the extent that deliberation is focused on participants' beliefs, we might ask whether it increases participants' knowledge (i.e., whether it creates more beliefs based on fact rather than on fiction or opinion). Common sense suggests that deliberation can sometimes have this effect, but it is not clear that group deliberation is better for this purpose than individual study in isolation. Experiments in public deliberation both online and offline suggest that group deliberation does not improve knowledge acquisition over isolated study of briefing materials, and that attitude change may come primarily from the briefing materials and not from the influence of the group. (Though as Peter Muhlberger points out, "deliberation may still be needed to motivate people to read the materials.")[35]

The available literature on outcome quality focuses mostly on cases in which a social scientist can plausibly pass judgment on the quality of a group's decisions. But of course, often deliberation is needed most in cases where there is no objectively best decision or standard for what constitutes a good idea. In such cases, democratic deliberation may be needed for a decision to have perceived legitimacy.[36] Contemporary design teams utilize techniques to encourage novelty in order to improve group brainstorming (e.g., guidelines like "withhold judgment" and "encourage wacky ideas").[37] We are unaware of rigorous studies testing whether these techniques can overcome the social pressures that tend to discourage novel thinking in online brainstorming sessions, but there is some evidence that being able to see the contributions of others in an online setting improves brainstorming.[38]

---

30 Brown (2000: 176), cited in Sunstein (2006: 60).

31 Sunstein (2006: 60).

32 Whyte (1993), cited in Myers (2008: 278). For a case study of bias in real world group decision-making, see Whyte and Levi (1994).

33 Myers (2008: 278-284). For refinements and challenges to group polarization results in the context of deliberation online versus face-to-face, see Muhlberger (2003, 2005b).

34 For a discussion of the literature, see Sunstein (2006: 60-64).

35 Private communication. See also Muhlberger and Weber (2006).

36 See Tyler and Lind (2001) on "procedural justice."

37 Salustri (2005). An influential set of principles is the four rules for brainstorming proposed by Osborn (1957), cited in Sutton and Hargadon (1996: 685): "don't criticize, quantity is wanted, combine and improve suggested ideas, and say all ideas that come to mind, no matter how wild."

*Collectivity*

As with the outcome dimension, whether a deliberation should result in a collective product is often dictated by context. To the extent that a deliberation designer has a choice, the available literature has shown more benefits to online deliberation for *individuals* who participate in online groups than it has for *group* online outcomes. In both online Deliberative Polls and scheduled political discussion online, individuals have been shown to have their attitudes affected by online discussion, and to evaluate the experience positively.[39] In contrast, research on online group decision-making has yet to demonstrate consistent benefits. A meta-analysis published in 2002 concluded that the extant literature at that time generally showed worse objective and subjective outcomes for online versus face-to-face group decision-making.[40] Since then, research comparing face-to-face and online decision-making in groups supports a more nuanced view that places emphasis on the capacity of a medium to support the type of communication between group members that is necessary for them to make a decision or solve a problem, given the skills and attitudes of group members in relation to the medium.[41]

As noted above, the literature comparing group deliberation to statistical (non-deliberating) sets of individuals shows mixed results for decision-making, primarily because groups can amplify biases. None of this amounts to a direct comparison of group versus individual outcomes as the goal of deliberation, but it does suggest, at least with the technology studied thus far, that online deliberation may have more benefits for individual learning and attitude change than for group-agreed outcomes, with the latter showing lesser or at best at par results in most online versus face-to-face studies. The latter results are limited by the quality of the technologies studied so far; however, they nevertheless provide evidence that group outcomes online show improvement as technology and the prevalence of skills to use it improve. It is also important to note that negative findings regarding group deliberation generally do not test deliberation structuring methods that have been developed to attenuate group biases.

**Population**

*Recruitment*

To our knowledge, no studies have systematically compared random versus nonrandom sampling for recruitment of participants in online deliberative forums. However, one of the arguments for random sampling is that if participants self-select, they will be less heterogeneous than the population as a whole (see chapters 3 and 5). One study looked at registrants and participation in a national online dialogue organized by the U.S. Environmental Protection Agency in 2000. Although participants were unrepresentative of the population as a whole in being heavy Internet users (and on related demographic measures), they were "representative of the broader public in the sense of bringing a diverse set of interest affiliations, attitudes about EPA, and geographical locations."[42]

---

[38] Michinov and Primois (2005).
[39] See Fishkin (2009c), Price (2009).
[40] Baltes, Dickson, Sherman, Bauer, and LaGanke (2002).
[41] Li (2007).
[42] Beierle (2004: 159).

A growing literature has looked at heterogeneity of self-selected groups in online forums. In Usenet groups, which tend to be focused on topics rather than defined by ideology, one study found that participants tend to be ideologically diverse and to reply more to opponents than to the like-minded.[43] On the Web, by contrast, there appear to be more pressures toward homophily (i.e., the tendency to connect with the like-minded), with some countervailing drivers of exposure to opposite viewpoints.[44] This may illustrate the sensitivity of Internet effects to specific technologies. Usenet forums historically have been organized by topic, bringing opposite sides together, while websites have loosely tended to reflect and serve particular points of view. In line with standard scientific practice, the designer of a deliberative forum should therefore be wary of selection biases in recruitment, and turn to random sampling or, less ideally, other techniques geared toward representative samples when inferences are to be drawn from the sample about a population, whether the deliberation is online or face-to-face.

### *Audience*

The question of whether the audience for a public-interest deliberation exercise should be public or private has not, to our knowledge, been studied in an online context. But the results of Deliberative Polls that have been broadcast nationally in the United Kingdom and United States have shown large opinion change effects on participants similar to those found in polls that were not broadcast or recorded for public viewing.[45] This suggests that a public audience, at the least, does not inhibit participants from changing their views in a group deliberation exercise. Work on "accountability" by psychologists provides evidence that an audience whose overall viewpoint is unknown to those being watched can encourage people to consider countervailing arguments more seriously, justify their own views more thoroughly, and qualify their opinions.[46] Theoretical arguments have, however, questioned whether politicians are likely to be as forthcoming when they have expert knowledge the public lacks, if their deliberations are conducted in public.[47] Also, a passive audience is likely to follow and remember deliberation less well than active participants.[48]

## Spatiotemporal Distance

### *Colocation*

A number of studies have specifically compared online with face-to-face deliberation. In this context, "online" usually means a form of deliberation that involves some sort of telecommunication, generally over the Internet. One such set of studies has looked at Deliberative Polling®.[49] Deliberative Polls (DPs) have been conducted online using synchronous voice discussion combined with an interface that shows each talking participant on the "chat line." The results are reported to be "broadly similar" to those of face-to-face Deliberative Polls, although more modest in their effects. This last fact has been attributed to the reduced personal involvement of individuals as they participate;. However, it .has been speculated that a longer lasting online DP might well produce opinion shifts as large as those in face-to-face DPs.[50]

---

43 Kelly, Fisher, and Smith (2009).

44 Lev-On and Manin (2009).

45 See Fishkin (1996, 2006a).

46 Lerner and Tetlock (1999: 256-257). Thanks to Peter Muhlberger for pointing us to this paper.

47 Stasavage (2004).

48 Schober and Clark (1989).

49 Luskin, Fishkin, and Iyengar (2006); Fishkin (2009c).

50 Fishkin (2009c: 31)

Another direct comparison of online versus face-to-face deliberation involved text-based instead of voice chat for the online condition. Online and face-to-face groups were randomly assigned, as was a control group. The deliberation was moderated and structured similarly in the two non-control conditions. This study found that both the online and face-to-face deliberations improved knowledge, efficacy, and political participation relative to the control group, and in about the same amounts. Opinion change did not show a similar pattern across the two groups, leading in opposite directions before and after, but the difference was not statistically significant.[51]

When combined, these studies suggest that online deliberation can produce outcomes that are similar to face-to-face deliberation, but that the level of involvement matters in the online case. Text-based chat may be less engaging than voice, and therefore less similar to face-to-face deliberation. Moreover, because the opinion shift in the face-to-face condition in the text-online study was not significant, the results indicate that at least in this exercise, online DPs are less effective than officially structured DPs. Before such broad-brush conclusions can be drawn, however, more research is needed.

One of the most systematic comparisons of online and face-to-face deliberation looked at the effects that face-to-face, synchronous online, and asynchronous online (over a 24-hour period) deliberations, respectively, had on participants' deliberative behavior within each condition. The study found that online deliberators expressed a greater variety of viewpoints (less conformity to a group's most popular opinion) and showed more equality of participation than their face-to-face counterparts, but that face-to-face deliberation was of higher quality, more likely to refer to personal experiences, and more enjoyable for participants.[52] Other research has also found that computer-mediated communication (CMC) reduces individual conformity to the dominant opinion of a group, relative to face-to-face deliberation in deliberative judgment tasks.[53] But although online deliberation, particularly of the asynchronous variety (see below) tends to promote equal participation relative to face-to-face deliberation, it does not appear to reduce the aversion to conflict that characterizes people who do not like to deliberate.[54]

A large and growing literature compares face-to-face and computer-mediated communication more generally. Among the most interesting findings has been a study of group problem solving in the Lost at Sea task.[55] CMC participants were "significantly better judges of whether they made a successful decision and agree[d] on the success of their group's decision to a greater extent than [face-to-face] group members."[56] Recalling the lack of general support in the extant literature for superior decisions in CMC settings discussed above, this suggests nonetheless that the online context may attenuate the social biases and distractions that are more likely to be manifested in face-to-face settings and that can interfere with realistic judgments of group effectiveness.[57]

---

51 Min (2007).

52 Tucey (2010).

53 King, Hartzel, Schilhavy, Melone, and McGuire (2010).

54 Neblo, Esterling, Kennedy, Lazer, and Sokhey (2010: 574).

55 In this task, participants must collectively rank 15 items in order of importance for survival from yachting accident, and success is measured against a consensus of experts (Roch and Ayman 2005: 20).

56 Roch and Ayman (2005: 29).

57 Roch and Ayman (2005: 28).

Another important issue for online communication is the extent to which users are satisfied with it relative to a face-to-face alternative. This is obviously quite dependent on the population of participants and the technology in question. CMC versus face-to-face studies of decision-making and problem solving have generally found that CMC can be as satisfying as face-to-face communication for groups that are not anonymous and that have sufficient time to complete the task (which tends to take longer in CMC in these studies than face-to-face).[58] In a couple of studies from neighboring fields, students were reported to consider synchronous voice communication to be the next best thing after face-to-face communication.[59]

A useful framework for describing results of online versus face-to-face studies of deliberation and decision-making is the set of theories that have been referred to in the group processes literature as "media capacity theories." The shared features of these theories have been described as follows:

> CMC is a more restricted medium than FTF [face-to-face] communication; hence whether CMC groups are more effective than FTF groups depends a great deal on the complexity of their tasks. When the task is complex, FTF groups should perform much better than CMC groups because their members need rich media to engage in a great deal of coordination, persuasion, and expression of opinions in their group interactions. In contrast, when the task is simple, CMC should be a sufficient channel for groups to accomplish their tasks, and thus CMC groups will perform just as well as FTF groups.[60]

Two sets of findings support this pattern: 1) CMC groups can take 4-10 times longer than face-to-face groups to finish assigned tasks due to the lack of nonverbal clues that aid understanding (grounding); and 2) CMC groups have often been found to have lower efficacy than face-to-face groups when performing tasks that require a lot of communication.[61]

*Cotemporality*

A comparison of online discussion forums initiated by the city of Hoogeveen in the Netherlands in 2001-2002 provides some data on the synchronous versus asynchronous forum question.[62] On a number of measures, the asynchronous text forum showed similar patterns to two synchronous text forums (e.g., special treatment based on status occurred in both kinds of forums). However, the asynchronous forum generated much longer postings (179 words on average compared with 30 and 46 in the synchronous forums). Synchronous deliberation seemed to score modestly higher on grounding (i.e., "mutual understanding"). The asynchronous forum generated more deliberation (replies), but also more verbal attacks. Because these results are based on observations rather than a controlled experiment, however, it is possible that they can be accounted for by population differences (or selection bias) in the forums.

A more controlled comparison was recently done between synchronous and 24-hour asynchronous forum deliberation about global warming and stem cell issues. While the asynchronous forum held over a 24 hour period produced more equality of participation than the synchronous forum, this experiment's results were consistent with those of the Hoogeveen study in that discussion was less interactive in the asynchronous than in the synchronous online forum – with more repetitive postings and position statements rather than dialogue, indicating lower grounding, in the former.[63] It has also been noted that "synchronous deliberation gives facilitators control over who can speak and also permits them to monitor and guide the nature and flow of the deliberative conversations"[64] relative to asynchronous deliberation.

---

58 See the literature reviewed in Li (2007: 598).
59 Brannon and Essex (2001); Lightner (2007).
60  Li (2007: 597).
61  Li (2007: 597-598).
62 Jankowski and van Os (2004).

At least a few studies in related fields have looked at the effects of the cotemporality variable in general for computer-mediated communication, leading to pertinent findings. For example, asynchronous text usually generates the most significant information exchanges (barring, presumably, face-to-face communication),[65] and synchronous text communication tends to be selected more often than asynchronous text for "community-building" efforts, although it can exclude unskilled typists to some degree.[66]

**Communication Medium**

*Modality*

Studies of the effects of different modalities that have focused specifically on online public deliberation have, to our knowledge, just looked at the relative effects of print versus online commenting on proposed regulations (e-rulemaking). None of the studies appear to have found large reliable differences in the quality or quantity of comments between these two media.[67] An interesting effect bearing on modality choice emerged from early structured deliberation experiments on political issues in the United States at the University of Pennsylvania. Text chat deliberators appeared more likely to show a pattern of more equal levels of discussion contributions across participants, relative to face-to-face voice dialogue, and this modality appeared to draw out those with minority opinions more than in live spoken dialogue.[68] This result in a controlled setting differs from findings from organically occurring online forums in which women and African-Americans were found to participate at disproportionately low levels.[69]

A large number of studies from computer mediated communication and human-computer interaction bear on the choice of modality for a deliberative forum. The optimal modality for each communication situation is sensitive to many variables. Other things being equal, however, the available research supports the idea that people both prefer and are more productive when they are speaking rather than writing, probably because speech is less cognitively demanding than writing,[70] but that people who are high in literacy prefer and absorb more information per unit time when they are reading text rather than listening to speech.[71] This suggests a role for the developing technology of automatic speech recognition (ASR). If software can efficiently translate spoken words into text, then the users of an online system may be able to interact more optimally.

---

63 Tucey (2010).

64 American Institutes for Research (2011: 131), citing Siu (2008).

65 Brannon and Essex (2001); Lightner (2007).

66 Brannon and Essex (2001).

67 Stanley, Weare, and Musso (2004); Schlosberg, Zavestoski, and Shulman (2009).

68 Price (2009).

69 Stromer-Galley and Wichowski (2010: 174).

70 Gould (1978) and Kroll (1978), cited in Kraut, Galegher, Fish, and Chalfonte (1992: 403).

71 Le Bigot, Rouet, and Jamet (2007).

Much work in the study of computer mediated communication has revolved around what we referred to earlier as "media capacity theories." *Media richness theory*, one well-known media capacity theory, has generated considerable research over the last three decades.[72] In this theory, a key variable predicting which modality a group will choose is *equivocality,* or the extent to which the correct solution to a group's task is ambiguous.[73] One formulation of media richness theory is the following:

> When equivocality is high, individuals are likely to have different interpretations of problems and may disagree as to what information is needed to shape a solution. These conditions require that individuals must first create a shared sense of the situation and then, through negotiation and feedback, formulate a common response. Daft and his colleagues argue that this requires a rich communication medium, one that, in our terminology, provides *interactivity* and *expressiveness*. A medium that provides interactivity permits communication partners to exchange information rapidly, adjusting their messages in response to signals of understanding or misunderstanding, questions, or interruptions [citation omitted]. A medium that permits expressiveness allows individuals to convey not only the content of their ideas but also intensity and subtleties of meaning through intonation, facial expression, or gestures. According to the contingency hypothesis, when task equivocality is high, media richness is essential to effective communication.[74]

Media richness theorists distinguish between "rich" and "lean" media, but this is usefully refined into the interactivity and expressiveness dimensions defined above. Modalities (text, speech, video, face-to-face) can be mapped onto expressiveness, and cotemporality can be mapped onto interactivity, thus defining a classification for two-way communication media under these two dimensions as shown in Table 6.2, with richness increasing as one moves from the upmost left toward the lowest right cell.

**Table 6.2: Two-Way Communication Media Representing Different Levels of Interactivity and Expressiveness**

| | | Expressiveness (modality) | | |
|---|---|---|---|---|
| | | *Low (text)* | *Moderate (speech)* | *High (video)* |
| **Interactivity (cotemporality)** | *Low (asynchronous)* | Email | Voice mail | Video mail |
| | *Moderate (on demand)* | Instant messaging (texting) | Instant voice messaging | Instant video messaging |
| | *High (synchronous)* | Synchronous text editing | Phone call (teleconferencing) | Video conferencing |

---

72 For example, Daft and Lengel (1986).

73 Kraut, Galegher, Fish, and Chalfonte (1992: 378).

74 Kraut, Galegher, Fish, and Chalfonte (1992: 378).

Both the interactivity/cotemporality and the expressiveness/modality dimensions correspond to tradeoffs for the users of media. Higher expressiveness (video, and to a lesser extent speech) enables participants to communicate and understand subtleties of sentiment, but at the cost of greater parallel processing requirements for the recipient, and therefore, potential failure to communicate intended information because of distraction and information overload. Lower expressiveness (text, then speech) enables message senders to control more precisely the information that is communicated, but with potential loss of emotional nuance of which the sender may be more aware than the receiver. Lower expressiveness therefore places a higher demand on communicators to wisely choose and interpret words. Higher interactivity (synchronous, and to a lesser extent on-demand) enables communicators to respond to each other more quickly, and therefore to avoid prolonged misunderstanding, but at the cost of greater cognitive demands on the sender, who must pay attention to the recipients' responses and reply to them under time pressure. Lower interactivity (asynchronous, and to a lesser extent on demand) gives both senders and receivers greater control over the pace at which they produce or understand a message, but with potential loss of communicative grounding. Additional important factors that may influence the choice of medium include whether communication is recorded and available for review later, and whether there is one, a few, or many recipients of a sender's message.

Media richness theory predicts that communicators will use rich media more when task equivocality (which may come from divergent preferences/emotions or from different beliefs/uncertainty) is high, and that they will use lean media more when task equivocality is low. Moreover, such theory suggests that measures such as quality, efficiency, efficacy, and satisfaction will improve for a communication if media richness and equivocality levels are well matched for a group's task. Many empirical studies have supported this theory to at least some extent, but the results overall have been mixed.[75] In a study of collaborative writing, for example, individuals who commented on a fictitious coauthor's manuscript "preferred text only for commenting on problems of spelling and grammar and preferred voice for commenting on missing the main point of a passage, on project status, and on the success of the collaboration." The latter tasks were interpreted to be higher in equivocality.[76] The presence of variables other than task equivocality, however, can interact with media richness. An alternative vein of media capacity theory emphasizes the social influence of people's skills, habits, and attitudes within a group in helping to determine which medium will be most effective for a given task.[77]

### *Emotivity*

A number of studies of computer-mediated communication have looked at the use and effects of emoticons in text communication. Emoticons can be included within plain text without further facilitation by the interface, but an interface can make available more visible and detailed emoticons for use in a message board. One study found that participants in a computer mediated decision-making session were more satisfied with their experience if emoticons were available to them.[78]

---

75 See for example, Dennis and Kinney (1998), Kahai and Cooper (2003).
76 Kraut, Galegher, Fish, and Chalfonte (1992: 395).
77 See for example, Fulk, Schmitz, and Steinfield (1990), cited in Li (2007: 597).
78 Rivera, Cooke, and Bauhs (1996).

We have already touched on evidence, under Structure and Cotemporality, that free-flowing asynchronous text forums are associated with less civil behavior. This may therefore make participants less comfortable. A study by social psychologists showed that email users tend to misinterpret the tone of messages even though they think they are interpreting them correctly, and that this can result in flame wars.[79]

As was discussed under Modality, the choice of a medium that supports speech, as opposed to text-only media, tends to enhance the ability to communicate emotion accurately. Speech also appears more likely to prevent emotional blow-ups. In a study of collaborative writing using different modalities, those using speech were more likely to use mitigating phrases to smooth over interactions, whereas text users were more likely to use blunt or "almost hostile" phrases.[80]

Computer mediated communication generally appears to be laden with emotion, with or without the use of emoticons. A review of the literature concluded that "there is no indication that CMC is a less emotional or less personally involving medium than F2F [face-to-face]. On the contrary, emotional communication online and offline is surprisingly similar, and if differences are found they show more frequent and explicit emotion communication in CMC than in F2F."[81]

*Fidelity*

A growing body of research has looked at the effect of transformed social interaction (TSI) on users of virtual reality systems. As yet, it is unclear what effects this would have on online deliberation per se, but evidence is suggestive. A general finding, replicated in several studies with avatars, is that transformations of facial images designed to increase a feeling of connection between participants (e.g., blending the face one is looking at with one's own through computerized morphing, maintaining eye contact that cannot happen naturally) are not detected as transformations, do not disrupt communication, and can lead to more satisfying, persuasive experiences for participants.[82]

**Deliberative Process**

*Facilitation*

Studies of online public deliberation have showed that online discussions moderated by government officials resulted in more respectful behavior by participants than is found in spontaneous, unofficial online discussion.[83] In a study of online deliberation organized by the government of the city of Hamburg, the researcher concluded that "the quality of debate was close to the rational-critical ideal of deliberative theory."[84] A few studies have looked systematically at the effect of moderators on online deliberative behavior. In an experimental study of Korean voters discussing an upcoming election, the presence of a moderator was found to significantly reduce the number of postings (pre-moderation), but improve deliberation quality and the likelihood that others' postings were read by a participant.[85] However, another study found that moderation involving censorship (i.e., the filtering of messages) can diminish trust in the forum.[86]

---

79 Epley and Kruger (2005). For a more positive assessment of the ability of a text forum (Usenet) to foster community, see Baym (1998).
80 Kraut, Galegher, Fish, and Chalfonte (1992: 399).
81 Derks, Fischer, and Bos (2008: 766).
82 Bailenson, Beall, Loomis, Blascovich, and Turk (2005).
83 Coleman (2004); Jensen (2003); Wright and Street (2007), all cited in Stromer-Galley and Wichowski (2010: 178) .
84 Albrecht (2006: 75).
85 Rhee and Kim (2009).

In a field experiment conducted in an online forum for discussing the future of New York's World Trade Center site, groups were given either "advanced" or "basic" facilitation, with the former involving professional facilitators who took a more active role in steering and summarizing discussions. Nonwhite (especially) and female residents were less likely to register for the discussions, but advanced facilitation appeared to boost participation for both groups relative to the basic condition, indicating that a more active approach might draw out underrepresented participants once they are part of the process.[87]

A growing volume of empirical study has focused on distributed moderation in systems such as Slashdot.org. Distributed moderation grew out of earlier work in "collaborative filtering"[88] that has found wide applications in electronic commerce, such as the Amazon.com system for suggesting items a user might like. Early analyses concluded that the Slashdot reputation and filtering system is relatively effective at identifying and directing users' attention to the highest value posts and comments, but with significant imperfections, such as biases created by earlier as opposed to later feedback.[89] More recent work has focused on making more efficient use of information in user behavior, for example by letting all users benefit from the extra effort that some users exert in modifying filter settings.[90]

### *Structure*

Although few studies have systematically manipulated structure in online public deliberation, a number of other studies of online forums provide insight into the structure question. For example, in the Hoogeveen, Netherlands study mentioned above, one of the two synchronous forums was much more restrictive in the postings it allowed.[91] This predictably led to fewer postings overall, and illustrates what may be a tradeoff between structure and volume of participation. At the same time, more structure has been shown in several studies to lead to more deliberative behavior.[92] Many forum designs build in deliberative structure, but a common finding, at least anecdotally, is that these structures can be oppressive and prevent people from interacting in the ways that they want.[93] Nonetheless, there appears to be a growing consensus among deliberation advocates that the open Internet is not conducive to deliberation. This sentiment has even been expressed by Jurgen Habermas himself:

> Use of the Internet has both broadened and fragmented the contexts of communication. This is why the Internet can have a subversive effect on intellectual life in authoritarian regimes. But at the same time, the less formal, horizontal cross-linking of communication channels weakens the achievements of traditional media. This focuses the attention of an anonymous and dispersed public on select topics and information, allowing citizens to concentrate on the same critically filtered issues and journalistic pieces at any given time. The price we pay for the growth in egalitarianism offered by the Internet is the decentralised access to unedited stories. In this medium, contributions by intellectuals lose their power to create a focus.[94]

---

86 Coleman, Hall, and Howell (2002); Wright (2009)

87 Trénel (2009).

88 Resnick, Iacovou, Suchak, Bergstrom, and Reidl (1994).

89 Lampe and Resnick (2004); Poor (2005).

90 Lampe, Johnston, and Resnick (2007).

91 Jankowski and van Os (2004).

92 See Janssen and Kies (2005) for an overview; see also Zhang (2005).

93 Schuler (2009).

Because "structure" is such a broad category of design possibilities, generic findings about structure are likely to be of less use to designers than are tests of particular techniques. Thus, this area is wide open for future research.

*Identifiability*

At least a few studies have looked specifically at the effect of forcing forum users to identify themselves. In a public deliberation context, Korean voters who were engaged in a controlled online discussion experiment about an upcoming election were found to be more engaged when allowed to post anonymously; however, identity cues were also found to improve externally rated discussion efficacy.[95] Consistent with these results, when a corporate online forum changed its policy so that employees could no longer post anonymously (measured by other participants' ability to see who they were), postings decreased significantly and became shorter, more narrowly focused, and less likely to generate a response.[96]

Early research on computer-mediated communication has generally found anonymity to boost CMC deliberation quality relative to face-to-face decision-making and problem solving. As mentioned earlier, a majority of studies have found CMC groups to perform less well in making decisions than face-to-face groups, but this difference disappears when CMC groups are anonymous.[97] Somewhat paradoxically, however, anonymous groups under CMC show less satisfaction than do face-to-face groups in decision tasks.[98] Anonymity can also lead to more irresponsible behavior, such as showing up late for the deliberation or being a "troublemaker."[99] The effects of identifiability may also be culturally dependent. For example, anonymity or reduced identifiability has been posited to overcome "Confucian norms of social hierarchy and status" present in East Asian societies.[100]

*Incentivization*

We have found only one study that systematically manipulated the presence and absence of rewards in an online deliberation forum – the Korean voter experiment – which split participants randomly into conditions where they received points that were "shown next to their login ID whenever they talked online." Points depended on their "frequency of postings, frequency of being read by someone else, and number of favorable replies." The points condition was found to produce more responses per message than the group without points. It also marginally improved a measure of "political discussion efficacy" that took account of "argument repertoire and other quality indices."[101]

---

94 Habermas (2006).

95 Rhee and Kim (2009).

96 Leshed (2009); cf. Lerner and Tetlock (1999).

97 Adams, Roch, and Ayman (2005); Baltes, Dickson, Sherman, Bauer, and LaGanke (2002); Becker-Beck, Wintermantel, and Borg (2005), all cited in Li (2007).

98 Li (2007: 598).

99 Tucey (2010: 23-24).

100 Min (2009: 454).

101 Rhee and Kim (2009: 227, 230).

**Remarks on Result Specificity and Culturalism**

The studies we have cited in this section, together with the design dimensions into which we have organized them, demonstrate that the main question facing future deliberation designers as they contemplate online designs will increasingly be not *whether* but *how* to use online tools. Early research in this area tended to treat technology as "a dichotomous variable … either present or not."[102] But as technologies have evolved, the range of experiences they offer now span a broad spectrum from the simple and truncated text of a Twitter message to the high verisimilitude of the most advanced virtual reality environments, as well as augmented reality systems that attempt to give us the best face-to-face and online experiences simultaneously. Online deliberation and its public version – online deliberative civic engagement – are rich areas of study and design because of the huge space of possibilities within them.

At the same time, we should guard against techno-determinism, or the tendency to draw strict conclusions about the effects of technology, even when the technology being studied is narrowly circumscribed and the results are robust thus far. Cultural practices can change in response to technology and other factors, so that what holds today might cease to be the case in ten years, or in a culture not yet studied.

One researcher has characterized the pair of principles described above as "a culturalist perspective on technology" which:

> means that we take into account the complexity of the social practices of usage as well as its symbolic dimension. It also means that we have to be sensitive to the specific technology used in a given case, as the corresponding practices differ greatly.[103]

**CONCLUSION**

The literature relevant to the design of online deliberation processes is vast, and the empirical research discussed above is better seen as a sample than as a comprehensive guide to results across the several dimensions we have covered, which are themselves a small subset of the choices facing a designer. There is a large volume of literature that suggests some guidelines for online deliberation design, but relatively few of the questions have been adequately answered. Choices often involve tradeoffs. Anonymity, for example, has been found to increase efficacy in decision-making for online groups, but at the cost of a less satisfying experience for participants.[104]

While the literature is complex, the general frameworks of media capacity theories and grounding provide a way to understand some of the findings we have discussed. In turn, these frameworks can assist online deliberation designers in making choices between different media or modalities. The environment for deliberation should be a good match for the needs and abilities of deliberators to resolve ambiguities and to achieve grounding, regardless of whether the task is to converge on a decision or just to hear each others' ideas. When a medium is insufficiently rich, or when participants are not able to use it effectively enough, then it gets in the way of success, and is likely to hurt the quality, efficiency, and efficacy of the deliberation, as well as subjective satisfaction with the process. This means, among other things, that the effectiveness of online deliberation can depend on specifics of the online medium that go beyond the choice dimensions we have discussed here, such as details of interface design and the level of richness afforded to communication. As technology is continually advancing, it is therefore difficult to induce reliable principles about online versus offline deliberation. Other sources of variation include differences across cultures[105] and time periods. Nevertheless, we believe that this literature has a few lessons to teach deliberative designers. A summary of four such lessons follows.

---

[102] Albrecht (2006: 75).

[103] Albrecht (2006: 75), who cite as an influence Suchman, Blomberg, Orr, and Trigg (1999).

[104] Li (2007: 598).

First, online forums can approach the impact of face-to-face deliberations if they are sufficiently engaging, media-rich enough for the deliberative task, or if the standard for success is individual attitude change. So far, voice deliberation in real-time seems to be more capable of achieving this than is asynchronous text deliberation.

Second, there appears to be a tradeoff between media that give people more time (asynchronous, text) and those that stress more direct engagement (synchronous, voice). The former appear better for encouraging participation, including by those less represented in live discussions, and they lead to longer contributions. But, on average, they may be less effective at fostering mutual understanding or changing people's minds. Before substantive conclusions can be drawn, however, much more research is needed.

Third, unstructured dialogue on the Internet does not seem to foster deliberative behavior by the standards of deliberation advocates. Structure and human facilitation appear to be needed for high quality deliberation in general, and can produce it when serious purpose is evident, as in the case of government-organized and facilitated online forums. Structure, moderation, and facilitation can also be provided in distributed ways, and can utilize software techniques that provide higher information value with less effort by users.

Fourth, allowing anonymous participation, even when users can be traced by administrators, appears to make people more willing to contribute to discussion, but it also lowers the sense of satisfaction participants feel, which is a common finding whenever a communication medium puts emotional distance between participants.

Applying theories such as media capacity, grounding, and the general findings of social and cognitive psychology suggests many questions and hypotheses that do not appear to have been well addressed in the literature we have seen. Examples of such questions include the following:

- What effects do different levels of media richness, or different modalities, have on deliberative outcomes? Do different media systematically produce different individual attitude changes or different group decisions?
- To what extent do characteristics of interfaces affect outcomes and the deliberative process? For example, does representing an individual participant visually in the context of others make participants more aware that the viewpoint of the focal participant is just one among many, and does this lead that viewpoint to be discounted relative to an interface that decontextualizes the focal participant?

---

[105] See Fung (2002), Kulikova and Perlmutter (2007), Robinson (2005), all cited in Stromer-Galley and Wichowski (2010: 180-181) and Min (2009).

- If there are effects found of the type mentioned above, what normative implications does this have for deliberation design?

The field of online deliberation is indeed young. We can look forward to learning much more as further studies are done, and as technology develops. For example, as speech recognition technology improves and becomes widely available, the advantages of online deliberation should increase, because users will be able to interact more flexibly and efficiently using both speech and text, switching between them as needed (e.g., speaking for production and reading for reception). At the same time, the necessary technology is not yet widely and easily available, and more research is needed to provide a compelling case that the classical goal of equivalent or better group decision-making is reliably attainable through the use of an online system.

Meanwhile, there are other reasons why a deliberation designer might choose an online forum over, or in addition to, a face-to-face one. The most obvious reasons have to do with cost, personal convenience for participants, and the opportunity to involve more participants who cannot be present for face-to-face deliberative civic engagement. And, in contexts where face-to-face meetings create barriers for some users due to shyness or other social dynamics such as prejudice, structured online dialogue may, even with the limitations of present technology, provide a superior and more equitable experience if the modality choices made by the designer are well matched to participants' skills and preferences.

Chapter 6                                                                                    Online Deliberation Design -  31